\documentclass[12pt]{article}

\usepackage{amsmath}
\usepackage{amssymb}

\usepackage{graphicx}

\begin{document}
\begin{flushright}
\hfill{ USTC-ICTS-08-21}\\
\end{flushright}
\vspace{4mm}

\begin{center}

{\Large \bf Reconstructing Quintom from Ricci Dark Energy}

\vspace{8mm}

{\large Chao-Jun Feng}

\vspace{5mm}

{\em
 Institute of Theoretical Physics, CAS,\\
 Beijing 100080, P.R.China\\
 Interdisciplinary Center for Theoretical Study, USTC,\\
 Hefei, Anhui 230026, P.R.China
 }\\
\bigskip
fengcj@itp.ac.cn
\end{center}

\vspace{7mm}

\noindent Holographic dark energy with Ricci scalar as IR cutoff called Ricci dark energy(RDE) probes the nature of
dark energy with respect to the holographic principle of quantum gravity theory. Scalar field dark energy models like
quintom are often regarded as an effective description of some underlying theory of dark energy. In this letter, we
find how the generalized ghost condensate model(GGC) that can easily realize quintom behavior can be used to
effectively describe the RDE and reconstruct the function $h(\phi)$ of GGC.

\newpage

\section{Introduction}
The accelerating cosmic expansion first inferred from the observations of distant type Ia supernovae
\cite{Riess:1998cb} has strongly confirmed by some other independent observations, such as the cosmic microwave
background radiation (CMBR) \cite{Spergel:2006hy} and Sloan Digital Sky Survey (SDSS) \cite{:2007wu}. An exotic form of
negative pressure matter called dark energy is used to explain this acceleration. The simplest candidate of dark energy
is the cosmological constant $\Lambda$, whose energy density remains constant during the evolution of the universe. The
cosmological model that consists of a mixture of the cosmological constant and cold dark matter is called LCDM model,
which provides an excellent explanation for the acceleration of the universe phenomenon and other existing
observational data. However, this model suffers the 'fine-tuning' problem and the 'cosmic coincidence' problem. To
alleviate or even solve these two problems, many dynamic dark energy models were proposed such as quintessence
mentioned, k-essence, tachyons, phantoms, ghost condensates and quintom etc.. Generically, we regard the scalar field
as an effective description of an underlying theory of dark energy, which we still do not known, because we do not
understand entirely the nature of dark energy before a complete theory of quantum gravity is established, since the
dark energy problem may be in principle a problem belongs to quantum gravity\cite{Witten:2000zk}.\\

Although we are lacking a quantum gravity theory today, we can still make some attempts to probe the nature of dark
energy according to some principle of quantum gravity. It is well known that the holographic principle is an important
result of the recent researches for exploring the quantum gravity(or string theory)\cite{holoprin}. So that the
holographic dark energy model (HDE) constructed in light of the holographic principle possesses some significant
features of an underlying theory of dark energy\cite{Zhang:2006av}. Recently, Gao et.al \cite{Gao:2007ep} have proposed
a holographic dark energy model in which the future event horizon is replaced by the inverse of the Ricci scalar
curvature, and they call this model the Ricci dark energy model(RDE). Of course, this model also respect the
holographic principle. For this model with proper parameters, the equation of state crosses $-1$, so it is a 'quintom'.
It has been shown in ref.\cite{general} that there is really a simple one-field model that can realize the quintom
model, called the generalized ghost condensate model(GGC). In this letter, we find some equivalence between the GGC and
RDE model, and we reconstruct the function $h(\phi)$ of GGC from RDE. In Section II, we will briefly review RDE model
and GGC model, and reconstruct the function $h(\phi)$ of GGC from RDE model in Section III. In the last section we will
give some conclusions.

\section{Briefly Review on RDE and GGC}

Holographic principle \cite{Bousso:2002ju} regards black holes as the maximally entropic objects of a given region, so
a self-consistent effective field theory with UV cutoff $\Lambda$ in a box of size $L$ should satisfy the Bekenstein
entropy bound \cite{Bekenstein:1973ur} $ (L\Lambda)^3\leq S_{BH}=\pi L^2M_{pl}^2 $, where $M_{pl}$ is the Planck mass ,
$S_{BH}$ is the entropy of black hole, and $L$ acts as an IR cutoff. Furthermore, Cohen et.al. \cite{Cohen:1998zx}
found that the total energy  should not exceed the mass of a black hole of the same size either, namely $
L^3\Lambda^4\leq LM_p^2 $. Under this assumption, Li \cite{Li:2004rb} proposed the holographic dark energy as follows
\begin{equation}\label{li}
    \rho_\Lambda = 3c^2M_p^2 L^{-2}
\end{equation}
where $c^2$ is a dimensionless constant. Since the holographic dark energy with Hubble horizon as its IR cutoff does
not give an accelerating universe \cite{Hsu:2004ri}, Li suggested to use the future event horizon instead of Hubble
horizon and particle horizon, then this model gives an accelerating universe and is consistent with current
observation\cite{Li:2004rb, Huang:2004ai}. For the recent works on holographic dark energy, see ref.
\cite{Zhang:2007sh,Sadjadi:2007ts, Saridakis:2007cy}. \\

Recently, Gao et.al \cite{Gao:2007ep} proposed the Ricci dark energy model(RDE), in which they take the Ricci scalar as
the IR cutoff. The Ricci scalar of FRW universe is given by $R = -6(\dot H + 2H^2 + k/a^2)$, where dot denotes a
derivative with respect to time $t$ and $k$ is the spatial curvature. The energy density of RDE is
\begin{equation}\label{Ricci DE}
    \rho_X = \frac{3\alpha}{8\pi G} \left(\dot H + 2H^2 + \frac{k}{a^2}\right) \propto R
\end{equation}
where the dimensionless coefficient $\alpha$ will be determined by observations. Solving the Friedmann equation they
find
\begin{equation}\label{Energy density of Ricci DE}
   \frac{8\pi G }{3H^2_0}\rho_X  = \frac{\alpha}{2-\alpha}\Omega_{m0}e^{-3x} + f_0e^{-(4-\frac{2}{\alpha})x}
\end{equation}
where $\Omega_{m0} \equiv 8\pi G\rho_{m0}/3H^2_0$, $x = \ln{a}$ and $f_0$ is an integration constant. Substituting the
expression of $\rho_X$ into the conservation equation of energy, namely $p_X = -\rho_X-\frac{1}{3}\frac{d\rho_X}{dx}$
they get the pressure of RDE:
\begin{equation}\label{pressure of X}
    p_X = -\frac{3H^2_0}{8\pi G }\left(\frac{2}{3\alpha}-\frac{1}{3}\right)f_0e^{-(4-\frac{2}{\alpha})x}
\end{equation}
Taking the observation values of parameters they find the $\alpha \simeq 0.46 $ and $f_0 \simeq 0.65$
\cite{Gao:2007ep}. The the equation of state of RDE at high redshifts is closed to zero and approaches $-1$ at present,
and in the future RDE will becomes a phantom. The energy density of RDE during big bang nucleosynthesis(BBN) is  much
smaller than that of other components of the universe ($\Omega_X |_{1MeV}<10^{-6}$ when $\alpha<1$), so it does
not affect BBN procedure.\\

The SN analysis using the Gold data\cite{Riess:1998cb} indicates that the parametrization of $H(z)$ which crosses the
cosmological-constant boundary ($w=-1$) shows a good fit to data. Consider the Lagrangian density of a general scalar
field $p(\phi, X)$, where $X=-(1/2)(\partial \phi)^2$ is the kinetic energy term. The energy density can be derived by
identifying the energy momentum tensor of the scalar field with that of a perfect fluid.
\begin{equation}\label{energy density of general scalar}
    \rho_{de} = 2Xp_X - p \,,
\end{equation}
where $p_X\equiv\partial p/\partial X$. Then, the dynamic equations for the scalar field in the flat FRW universe read
\begin{eqnarray}
   &H^2&=\frac{8\pi G}{3} \left(\rho_m +  2Xp_X - p \right) \label{eom1 of general scalar}\\
   &\dot H&=-4\pi G \left(\rho_m+2Xp_X\right)\label{eom2 of general scalar} \,,
\end{eqnarray}
where $X=\dot\phi^2/2$ in the cosmological context. Making use of the energy conservation law of matter $\dot\rho_m +
3H\rho_m = 0$, we find $\rho_m = \Omega_{m0}\rho_{c0}e^{-3x}$, where $\rho_{c0}\equiv3H^2_0/(8\pi G)$ represents the
present critical density of the universe. One can rewrite eq.(\ref{eom1 of general scalar}) and (\ref{eom2 of general
scalar}) by using a dimensionless quantity $ r \equiv H^2/H^2_0$ as follows
\begin{eqnarray}
  p &=& -\rho_{c0}\left( r + \frac{r'}{3} \right) \label{new eom1 of general scalar} \\
  \phi'^2p_X &=& -\frac{3}{8\pi G r} \left(\frac{r'}{3} + \Omega_{m0}e^{-3x}\right) \label{new eom2 of general scalar}
\end{eqnarray}
where prime denotes a derivative with respect to $x\equiv\ln a$. The equation of state of dark energy is given by
\begin{equation}\label{eos for general scalar}
    w = \frac{p}{\dot\phi^2p_X-p} = \frac{p}{r\rho_{c0}-\rho_m}= \frac{r+r'/3}{\Omega_{m0}e^{-3x}-r} \,.
\end{equation}
It should be noticed that if we establish a correspondence between RDE model and the scalar field dark energy, we
should choose a scalar field model, which can cross the cosmological-constant boundary  \cite{Zhang:2006qu}.

\section{Reconstructing GGC from RDE}

Consider the generalized ghost condensate model proposed in ref.\cite{general} with Lagrangian density
\begin{equation}\label{ggc lagrangian}
    p = -X + h(\phi)X^2 \,,
\end{equation}
where $h(\phi)$ is a function in terms of scalar field $\phi$, and $h(\phi)=ce^{\lambda\phi}$ corresponds to the
dilatonic ghost condensate model. From eq.(\ref{new eom1 of general scalar}) and (\ref{new eom2 of general scalar}), we
get the following dynamic equations for GGC
\begin{eqnarray}
  \phi'^2 &=& \frac{3}{8\pi G r}\left(4r+r'-\Omega_{m0}e^{-3x}\right) \label{equation1}\\
   h(\phi)&=& \frac{3}{(4\pi G)^2\phi'^4r^2}\left(-3r-r'+4\pi G r\phi'^2\right)\rho_{c0}^{-1} \label{equation2}\, .
\end{eqnarray}
In the following, we construct the correspondence between GGC and RDE, namely $\rho_{de} = \rho_X$,  which provids that
\begin{equation}\label{r value}
    r = \frac{H^2}{H^2_0} = \Omega_{m0}e^{-3x} + \frac{8\pi G}{3H_0^2}\rho_{de} \, .
\end{equation}
And by using eq.(\ref{Energy density of Ricci DE}), we get
\begin{equation}\label{r value result}
    r = \frac{2}{2-\alpha}\Omega_{m0}e^{-3x} + f_0e^{-(4-\frac{2}{\alpha})x}
\end{equation}
and
\begin{equation}\label{r prime value}
    r' = -\frac{6}{2-\alpha}\Omega_{m0}e^{-3x} -\left(4-\frac{2}{\alpha}\right) f_0e^{-(4-\frac{2}{\alpha})x} \,.
\end{equation}\\
Thus, we can numerically solve (\ref{equation1}) and (\ref{equation2}) to obtain the function $h(\phi)$, which plotted
in Fig.1, and the scalar field $\phi$ is also plotted in Fig.2.

\bigskip{
    \vbox{
            {
                \nobreak
                \centerline
                {
                    \includegraphics[scale=1.0]{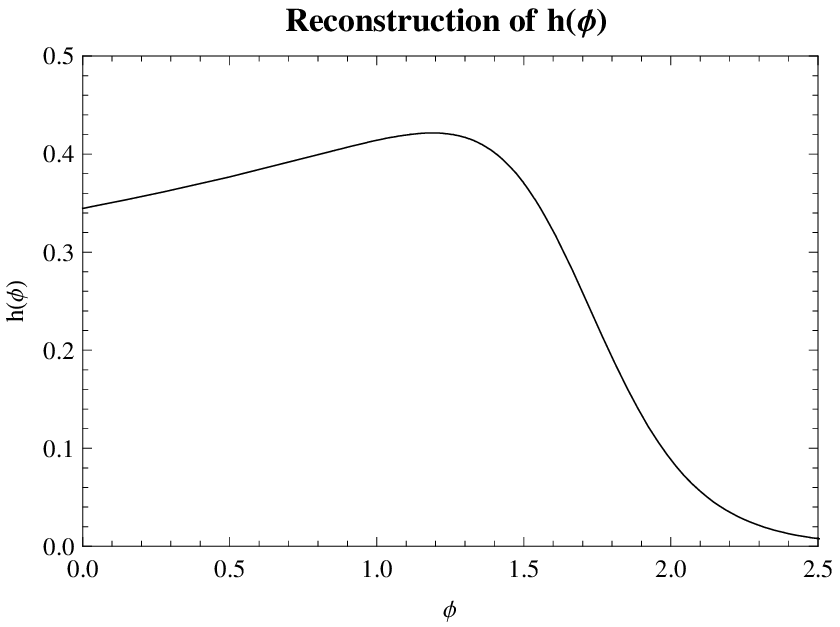}
                }
                \nobreak
                \bigskip
                {\raggedright\it \vbox
                    {
                        {\bf Figure 1.}
                        {\it Reconstruction of the generalized ghost model according to the Ricci dark energy model. The function $h(\phi)$ is plotted in unit of $\rho_{c0}^{-1}$ with $\phi$  in unit of $(4\pi
                        G)^{-1/2} = \sqrt{2}M_p$.
                        }
                    }
                }

            }
        }
\bigskip}

\bigskip{
    \vbox{
            {
                \nobreak
                \centerline
                {
                    \includegraphics[scale=1.0]{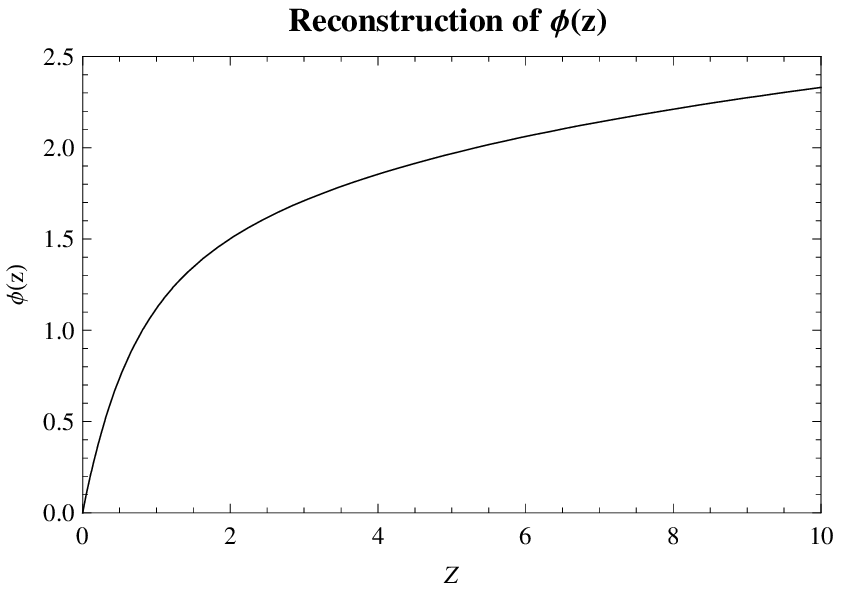}
                }
                \nobreak
                \bigskip
                {\raggedright\it \vbox
                    {
                        {\bf Figure 2.}
                        {\it Evolution of the scalar field $\phi(z)$ in unit of $\sqrt{2}M_p$ reconstructed according to the Ricci dark
                        energy.
                        }
                    }
                }

            }
        }
\bigskip}

From Fig.1 and Fig.2, one can see that $h(\phi)$ vanishes when $\phi$ is larger, which corresponding to the earlier
universe. From eq.(\ref{eos for general scalar}), we know that when $p_X = 0$, i.e. $hX = 1/2$, the equation of state
will cross the cosmological-constant boundary, and the universe enter the phantom region ($p_X < 0$) without
discontinuous behavior of $h$ and $X$, see Fig.3 and Fig.4 . Furthermore, our results  are consistent with that of
ref.\cite{general}, in which the authors have reconstructed the generalized ghost condensate model from the SN Gold
data\cite{cross1}. And the shapes of the curvatures seems much better than the reconstruction from HDE
\cite{Zhang:2006qu} to compare with that of ref\cite{general}.\\

\bigskip{
    \vbox{
            {
                \nobreak
                \centerline
                {
                    \includegraphics[scale=1.0]{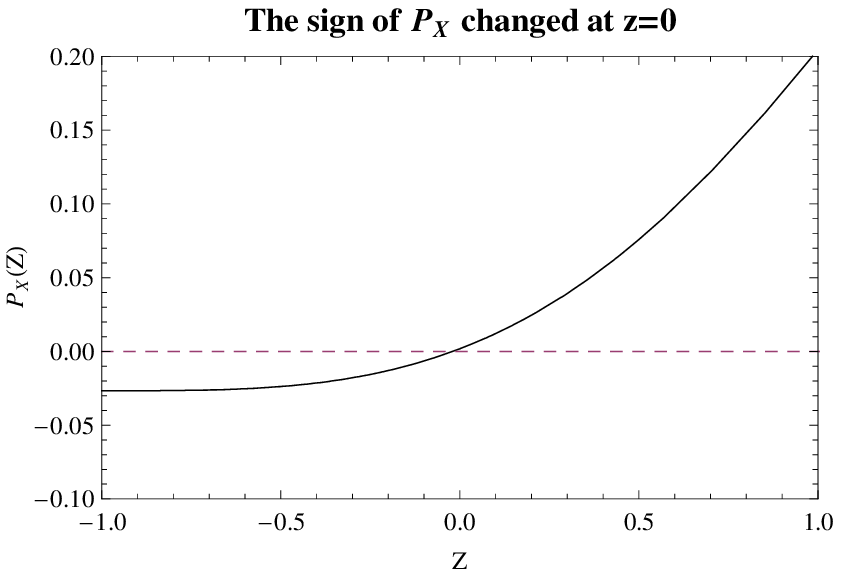}
                }
                \nobreak
                \bigskip
                {\raggedright\it \vbox
                    {
                        {\bf Figure 3.}
                        {\it The sign of $p_X = -1 + 2hX$ changed at $z=0$, and the universe enter the phantom region.
                        }
                    }
                }

            }
        }
\bigskip}

The reconstructed evolutions of equation of state $w$ in terms of $\phi$ are plotted in Fig.4.

\bigskip{
    \vbox{
            {
                \nobreak
                \centerline
                {
                    \includegraphics[scale=1.0]{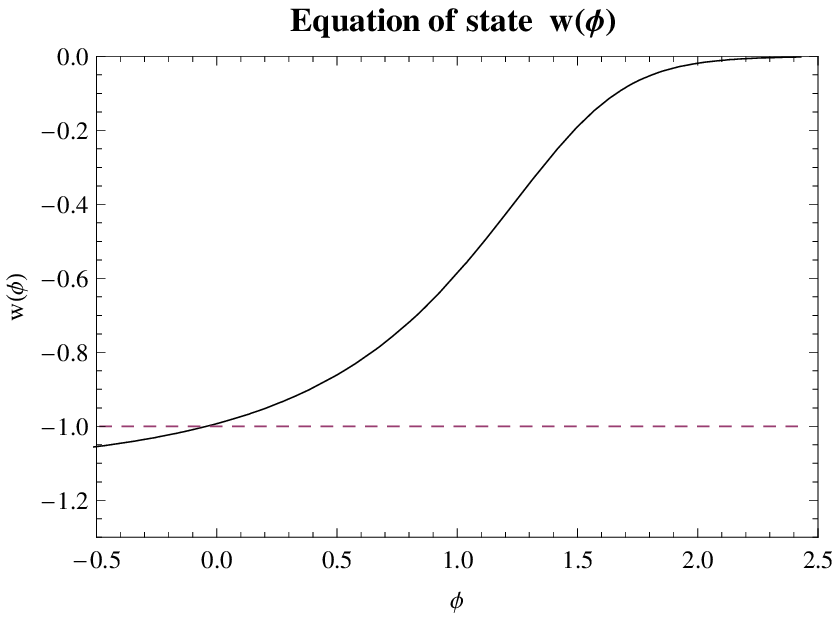}
                }
                \nobreak
                \bigskip
                {\raggedright\it \vbox
                    {
                        {\bf Figure 4.}
                        {\it Equation of state in term of $\phi$ for reconstructed GCC with $\alpha = 0.46$ and
                        $f_0=0.65$.
                        }
                    }
                }

            }
        }
\bigskip}

From Fig.4, one can see clearly that GGC mimic the evolving behavior of equation of state in RDE well. The equation of
state in GGC crosses $-1$ at $z=0$ and enter phantom region in the future, so it does realize a quintom-like dark
energy .

\section{Conclusions}

In conclusion, we adopt a correspondence between the Ricci dark energy model and a scalar field dark energy (GGC)
suggested in ref.\cite{Zhang:2006qu}. The current observation data imply RDE as quintom-like dark energy, i.e. the
equation of state of dark energy crosses the cosmological-constant boundary $w=-1$ during the evolution of the
universe. If the scalar field theory is regarded as an effective description of the dark energy, we should be capable
of using the scalar field model to mimic the evolving behavior of RDE and reconstructing the scalar model. The
generalized ghost condensate models(GGC) is a good choice for such a reconstruction since it can realize the quintom
behavior easily. Thus, we reconstructed the function $h(\phi)$ of the GGC using parameters $\alpha\approx0.46$ and
$f_0\approx0.65$ in RDE given in ref.\cite{Gao:2007ep} . We find that our reconstruction results in Fig.1 and Fig.2 are
consistent with that of ref.\cite{general}, in which the authors have reconstructed GGC from the SN Gold
data\cite{cross1}. And the shapes of the curvatures in Fig.1 and Fig.2 seems much better than the reconstruction from
HDE \cite{Zhang:2006qu} to compare with that of ref.\cite{general}. We also hope that the future high precision
observation data may be able to determine the parameters of RDE and reveal some significant features of the underlying
theory of dark energy.

\section*{ACKNOWLEDGEMENTS}
The author would like to thank Miao Li for a careful reading of the manuscript and valuable suggestions. We would like
to thank Xin Zhang for kind help in programme.\\



\begin{thebibliography}{99}
\bibitem{Riess:1998cb}
  A.~G.~Riess {\it et al.}  [Supernova Search Team Collaboration],
   ``Observational Evidence from Supernovae for an Accelerating Universe and a
  Astron.\ J.\  {\bf 116}, 1009 (1998)
  [arXiv:astro-ph/9805201].

  S.~Perlmutter {\it et al.}  [Supernova Cosmology Project Collaboration],
  Astrophys.\ J.\  {\bf 517}, 565 (1999)
  [arXiv:astro-ph/9812133].

\bibitem{Spergel:2006hy}
  D.~N.~Spergel {\it et al.}  [WMAP Collaboration],
   ``Wilkinson Microwave Anisotropy Probe (WMAP) three year results:
  Astrophys.\ J.\ Suppl.\  {\bf 170}, 377 (2007)
  [arXiv:astro-ph/0603449].

  E.~Komatsu {\it et al.}  [WMAP Collaboration],
  arXiv:0803.0547 [astro-ph].

\bibitem{:2007wu}
  J.~K.~Adelman-McCarthy {\it et al.}  [SDSS Collaboration],
  arXiv:0707.3413 [astro-ph].

\bibitem{Witten:2000zk}
  E.~Witten,
  hep-ph/0002297.

\bibitem{holoprin}
  G.~'t Hooft,
  gr-qc/9310026;\\
  L.~Susskind,
  J.\ Math.\ Phys.\  {\bf 36}, 6377 (1995)
  [hep-th/9409089].

\bibitem{Zhang:2006av}
  X.~Zhang,
  Phys.\ Lett.\  B {\bf 648}, 1 (2007)
  [arXiv:astro-ph/0604484].



\bibitem{Gao:2007ep}
  C.~Gao, X.~Chen and Y.~G.~Shen,
  [arXiv:0712.1394 astro-ph].


\bibitem{general}
S. Tsujikawa, Phys. Rev. D {\bf72}, 083512 (2005)
[astro-ph/0508542];\\
E. J. Copeland, M. Sami and S. Tsujikawa, hep-th/0603057.


\bibitem{Bousso:2002ju}
  R.~Bousso,
  Rev.\ Mod.\ Phys.\  {\bf 74}, 825 (2002)
  [arXiv:hep-th/0203101].

\bibitem{Bekenstein:1973ur}
  J.~D.~Bekenstein,
  Phys.\ Rev.\  D {\bf 7}, 2333 (1973).

  J.~D.~Bekenstein,
   ``A Universal Upper Bound On The Entropy To Energy Ratio For Bounded
  Phys.\ Rev.\  D {\bf 23}, 287 (1981).

\bibitem{Cohen:1998zx}
  A.~G.~Cohen, D.~B.~Kaplan and A.~E.~Nelson,
  Phys.\ Rev.\ Lett.\  {\bf 82}, 4971 (1999)
  [arXiv:hep-th/9803132].

\bibitem{Li:2004rb}
  M.~Li,
  Phys.\ Lett.\  B {\bf 603}, 1 (2004)
  [arXiv:hep-th/0403127].

\bibitem{Hsu:2004ri}
  S.~D.~H.~Hsu,
  Phys.\ Lett.\  B {\bf 594}, 13 (2004)
  [arXiv:hep-th/0403052].

\bibitem{Huang:2004ai}
  Q.~G.~Huang and M.~Li,
  JCAP {\bf 0408}, 013 (2004)
  [arXiv:astro-ph/0404229].

  Q.~G.~Huang and M.~Li,
  JCAP {\bf 0503}, 001 (2005)
  [arXiv:hep-th/0410095].


\bibitem{Zhang:2007sh}
  X.~Zhang and F.~Q.~Wu,
  Phys.\ Rev.\  D {\bf 76}, 023502 (2007)
  [arXiv:astro-ph/0701405].

  Y.~S.~Myung,
  Phys.\ Lett.\  B {\bf 649}, 247 (2007)
  [arXiv:gr-qc/0702032].

  H.~b.~Zhang, W.~Zhong, Z.~H.~Zhu and S.~He,
  Phys.\ Rev.\  D {\bf 76}, 123508 (2007)
  [arXiv:0705.4409 astro-ph].

  Z.~Y.~Sun and Y.~G.~Shen,
  Int.\ J.\ Theor.\ Phys.\  {\bf 46}, 877 (2007).

  Y.~S.~Myung,
  Phys.\ Lett.\  B {\bf 652}, 223 (2007)
  [arXiv:0706.3757 gr-qc].

  C.~Feng, B.~Wang, Y.~Gong and R.~K.~Su,
  JCAP {\bf 0709}, 005 (2007)
  [arXiv:0706.4033 astro-ph].

  H.~Wei and S.~N.~Zhang,
  Phys.\ Rev.\  D {\bf 76}, 063003 (2007)
  [arXiv:0707.2129 astro-ph].

  B.~Guberina,
  [arXiv:0707.3778 gr-qc].

  B.~C.~Paul, P.~Thakur and A.~Saha,
  [arXiv:0707.4625 gr-qc].

  J.~f.~Zhang, X.~Zhang and H.~y.~Liu,
  Eur.\ Phys.\ J.\  C {\bf 52}, 693 (2007)
  [arXiv:0708.3121 hep-th].

  C.~J.~Feng,
  [arXiv:0709.2456 hep-th].

  Y.~Z.~Ma and Y.~Gong,
  [arXiv:0711.1641 astro-ph].

  R.~Horvat,
  [arXiv:0711.4013 gr-qc].

\bibitem{Sadjadi:2007ts}
  H.~M.~Sadjadi,
  JCAP {\bf 0702}, 026 (2007)
  [arXiv:gr-qc/0701074].

  M.~R.~Setare,
  JCAP {\bf 0701}, 023 (2007)
  [arXiv:hep-th/0701242].

  M.~R.~Setare and E.~C.~Vagenas,
  [arXiv:0704.2070 hep-th].

  J.~Zhang, X.~Zhang and H.~Liu,
  Phys.\ Lett.\  B {\bf 659}, 26 (2008)
  [arXiv:0705.4145 astro-ph].

  Q.~Wu, Y.~Gong, A.~Wang and J.~S.~Alcaniz,
  Phys.\ Lett.\  B {\bf 659}, 34 (2008)
  [arXiv:0705.1006 astro-ph].

  K.~Y.~Kim, H.~W.~Lee and Y.~S.~Myung,
  Mod.\ Phys.\ Lett.\  A {\bf 22}, 2631 (2007)
  [arXiv:0706.2444 gr-qc].

  M.~R.~Setare,
  [arXiv:0708.3284 hep-th].

\bibitem{Saridakis:2007cy}
  E.~N.~Saridakis,
  Phys.\ Lett.\  B {\bf 660}, 138 (2008)
  [arXiv:0712.2228 hep-th].

  E.~N.~Saridakis,
  JCAP {\bf 0804}, 020 (2008)
  [arXiv:0712.2672 astro-ph].

  X.~Wu and Z.~H.~Zhu,
  Phys.\ Lett.\  B {\bf 660}, 293 (2008)
  [arXiv:0712.3603 astro-ph].

  E.~N.~Saridakis,
  Phys.\ Lett.\  B {\bf 661}, 335 (2008)
  [arXiv:0712.3806 [gr-qc].

  A.~A.~Sen and D.~Pavon,
  [arXiv:0801.0280 astro-ph].

  M.~Li, C.~Lin and Y.~Wang,
  [arXiv:0801.1407 astro-ph].

  K.~Karwan,
  JCAP {\bf 0805}, 011 (2008)
  [arXiv:0801.1755 astro-ph].

  M.~R.~Setare and E.~C.~Vagenas,
  [arXiv:0801.4478 hep-th].

  A.~J.~M.~Medved,
  [arXiv:0802.1753 hep-th].

  Y.~S.~Myung and M.~G.~Seo,
  [arXiv:0803.2913 gr-qc].

  B.~Nayak and L.~P.~Singh,
  arXiv:0803.2930 [gr-qc].

  L.~Xu and J.~Lu,
  arXiv:0804.2925 [astro-ph].

  K.~Y.~Kim, H.~W.~Lee and Y.~S.~Myung,
  arXiv:0805.3941 [gr-qc].

  C.~J.~Feng,
  arXiv:0806.0673 [hep-th].

  H.~Mohseni Sadjadi and N.~Vadood,
  arXiv:0806.2767 [gr-qc].

  R.~Horvat,
  arXiv:0806.4825 [gr-qc].

  N.~Cruz, S.~Lepe, F.~Pena and J.~Saavedra,
  arXiv:0807.3854 [gr-qc].

  Y.~Bisabr,
  arXiv:0808.1424 [gr-qc].

  C.~J.~Feng,
  arXiv:0809.2502 [hep-th].

  B.~C.~Paul, P.~Thakur and A.~Saha,
  arXiv:0809.3491 [hep-th].

  M.~R.~Setare and E.~N.~Saridakis,
  arXiv:0810.0645 [hep-th].


\bibitem{Zhang:2006qu}
  X.~Zhang,
  Phys.\ Rev.\  D {\bf 74}, 103505 (2006)
  [arXiv:astro-ph/0609699].

\bibitem{cross1}
  U.~Alam, V.~Sahni and A.~A.~Starobinsky,
  JCAP {\bf 0406}, 008 (2004)
  [astro-ph/0403687].


\end{thebibliography}
\end{document}